\documentclass[aps,showpacs,preprintnumbers,nofootinbib]{revtex4}
\usepackage{amssymb}
\usepackage{amsfonts}
\usepackage{amsmath}
\usepackage{graphicx}
\usepackage{dcolumn}
\usepackage{bm}
\usepackage[latin1]{inputenc}
\usepackage[dvips]{hyperref}
\usepackage{graphicx}

\def\be{\begin{equation}}
\def\ee{\end{equation}}
\def\bea{\begin{eqnarray}}
\def\eea{\end{eqnarray}}

\begin{document}

\title{Dirac Quasinormal Modes of New Type Black Holes in New Massive Gravity}
\author{P. A. Gonz\'{a}lez}
\email{pablo.gonzalez@udp.cl}
\affiliation{Facultad de Ingenier\'{\i}a, Universidad Diego Portales, Avenida Ej\'{e}%
rcito Libertador 441, Casilla 298-V, Santiago, Chile.}
\author{Y. V\'{a}squez.}
\email{yerko.vasquez@ufrontera.cl}
\affiliation{Departamento de F\'{\i}sica, Facultad de Ciencias, Universidad de La Serena,\\ 
Avenida Cisternas 1200, La Serena, Chile.}
\date{\today }

\begin{abstract}
We study new type black holes in three-dimensional New Massive Gravity and
we calculate analytically the quasinormal modes for fermionic perturbations
for some special cases. Then, we show that for these cases the new type
black holes are stable under fermionic field perturbations.
\end{abstract}

\maketitle


\section{Introduction}

In recent years, there has been a remarkable interest in the study of
three-dimensional models of gravity. Apart of BTZ black hole \cite%
{Banados:1992wn}, which is a solution to the Einstein equations with a
negative cosmological constant, remarkable attention was addressed to
Topologically Massive Gravity (TMG), which is a generalization of
three-dimensional GR that amounts to augment the Einstein-Hilbert action
adding a Chern-Simons gravitational term, \cite{Deser:1981wh}. Here, the
propagating degree of freedom is a massive graviton. TMG also admits the BTZ
(and other) black holes as exact solutions. The renewed interest on TMG
relies on the possibility of constructing a chiral theory of gravity at a
special point of the space of parameters, \cite{Deser:1982vy}.

On the other hand, Bergshoeff, Hohm and Townsend (BHT) introduced another
three-dimensional massive gravity theory, which is known as New Massive Gravity
(NMG), being the action the standard Einstein-Hilbert term with a specific
combination of scalar curvature square term and Ricci tensor square one \cite%
{Bergshoeff:2009hq, Bergshoeff:2009aq, Bergshoeff:2009tb, Bergshoeff:2009fj,
Andringa:2009yc, Bergshoeff:2010mf}, and it is equivalent at the linearized
level to the (unitary) Fierz-Pauli action for a massive spin-2 field \cite%
{Bergshoeff:2009hq}. The model in three dimensions is indeed unitary in the
tree-level, but the corresponding model in higher dimensions is not so due
to the appearance of non-unitary massless spin-2 modes \cite{Nakasone:2009bn}%
. Also, NMG admits warped AdS black holes \cite{Clement:2009gq}, AdS waves 
\cite{AyonBeato:2009yq, Clement:2009ka}, asymptotically Lifshitz black holes 
\cite{AyonBeato:2009nh}, gravitational solitons, kinks and wormholes \cite%
{Oliva:2009ip}, for further aspects of the BHT theory see \cite{Kim:2009jm,
Oda:2009ys, Liu:2009pha, Nakasone:2009vt, Deser:2009hb}. Besides, asymptotically AdS and Lifshitz black holes in NMG dressed by a (non)minimally coupled scalar field have been studied recently in \cite{Correa:2014ika}. It is worth mention
that TMG and NMG share common features, however there are different
aspects, one of them is the existence of the new type of black holes, for a
specific combination of parameters in the NMG Lagrangian, which was
discovered by BHT, and which are known as new type black holes.

The particular motivation of this work is to calculate the quasinormal modes
(QNMs) for fermionic field perturbations in the background of the new type
black holes in three-dimensional New Massive Gravity and study the stability
of these black holes under fermionic perturbations. The QNMs and their
quasinormal frequencies (QNFs) are an important property of black holes and
have a long history, \cite{Regge:1957td, Zerilli:1971wd, Zerilli:1970se,
Kokkotas:1999bd, Nollert:1999ji, Konoplya:2011qq}. It is known that the
presence of event horizons dampens the vibration modes of a matter field
that evolves perturbatively in the exterior region. In this way, the system
is intrinsically dissipative, i.e., there is no temporary symmetry. In
general, the oscillation frequencies are complex, therefore the system is
not Hermitian. Nevertheless, the oscillation frequency of these modes is
independent of the initial conditions and it only depends on the parameters
(mass, charge and angular momentum) and the fundamental constants (Newton
constant and cosmological constant) that describe a black hole just like the
parameters that define the test field. In three-dimensional spacetime, the QNMs of the BTZ black hole have been studied in \cite{Chan:1996yk}-\cite{Birmingham:2001pj}, and the QNMs for scalar field
perturbations in the background of new type black holes in NMG was studied
in \cite{Kwon:2011ey}.

The QNMs give information about the stability of black holes under matter
fields that evolves perturbatively in the exterior region of them, without
backreacting on the metric. Also, the QNMs determine how fast a thermal
state in the boundary theory will reach thermal equilibrium according to the
AdS/CFT correspondence \cite{Maldacena:1997re}, where the relaxation time of
a thermal state of the boundary thermal theory is proportional to the
inverse of the imaginary part of the QNMs of the dual gravity background 
\cite{Horowitz:1999jd}. In the context of black hole thermodynamics, the
QNMs allow the quantum area spectrum of the black hole horizon to be
studied, as well as the mass and the entropy spectrum. In this regard,
Bekenstein \cite{Bekenstein:1974jk} was the first to propose the idea that
in quantum gravity the area of black hole horizon is quantized, leading to a
discrete spectrum which is evenly spaced. Then, Hod \cite{Hod:1998vk}
conjectured that the asymptotic QNF is related to the quantized black hole
area, by identifying the vibrational frequency with the real part of the
QNFs. However, it is not universal for every black hole background. Then,
Kunstatter \cite{Kunstatter:2002pj} propose that the black hole spectrum can
be obtained by imposing the Bohr-Sommerfeld quantization condition to an
adiabatic invariant quantity involving the energy and the vibrational
frequency. Furthermore, Maggiore \cite{Maggiore:2007nq} argued that in the
large damping limit the identification of the vibrational frequency with the
imaginary part of the QNF could lead to the Bekenstein universal bound.
Then, the consequences of these proposals were studied in several
spacetimes, for instance, see \cite{Kwon:2011ey}, where the authors comment
on AdS/CFT correspondence and entropy/area spectrum for the new type black
holes. Besides, in \cite{Corda:2012tz}-\cite{Corda:2013paa} the authors discuss a connection between Hawking radiation and black hole quasinormal modes, which is important in the route to quantize gravity, because one can
naturally interpret black hole quasinormal modes in terms of quantum levels.

The paper is organized as follows. In Sec. II we give a brief review of the
new type black holes in three-dimensional New Massive Gravity. In Sec. III
we calculate the exact QNMs of fermionic perturbations for the new type black
holes. Finally, conclusions are
presented in Sec. IV.

\section{New Type Black Holes}

In this work we will consider a matter distribution outside the event
horizon of three-dimensional black holes, which are solutions of NMG and
conformal gravity in three dimensions \cite{Oliva:2009hz}. The action of NMG
theory is given by 
\begin{equation}
S=\frac{1}{16\pi G}\int d^{3}x\sqrt{-g}\left( R-2\lambda -\frac{1}{\mu ^{2}}%
K\right)~,
\end{equation}
where $K$ is given by the following quadratic terms of the curvature 
\begin{equation}
K=R_{\mu \nu }R^{\mu \nu }-\frac{3}{8}R^{2}~,
\end{equation}
where $R$ is the Ricci scalar, and $\lambda $ is the cosmological constant.
At the special case $\mu ^{2}=\lambda $ the theory admits the following metric
as solution 
\begin{equation}
ds^{2}=-f\left( r\right) dt^{2}+\frac{dr^{2}}{f\left( r\right) }+r^{2}d\phi
^{2}~,  \label{metric}
\end{equation}
where $f\left( r\right) =Ar^{2}+Br+C$. In the case that $B\neq 0$ there is a
curvature singularity due to the Ricci scalar diverges at $r=0$
\begin{equation}
R=-6A-\frac{2B}{r}~.
\end{equation}
The metric (\ref{metric}) is conformally flat, and hence solve the field
equations of conformal gravity in three dimensions too, as we mentioned. The
parameter $A$ is proportional to the cosmological constant $\lambda$, $B$ is
a kind of "gravitational hair" and $C$ is related to the mass of the
spacetime. Note that depending of the values of the parameters $A, B$ and $C$
this metric may represents a black hole or not, being the black holes
solutions known as new type black holes. The roots of $f\left( r\right) $
are
\begin{equation}
r_{\pm }=\frac{1}{2A}\left( -B\pm \sqrt{B^{2}-4AC}\right) ~,
\end{equation}
for $r_{+}>0$ the metric represents new type black holes with an event
horizon located at $r_{+}$. Also, in some cases may exist a cosmological
horizon $r_{-}$, with $r_{-}>r_{+}$. For $A>0$, the new type black holes are
asymptotically anti-de Sitter, for $A<0$ are asymptotically de Sitter and
for $A=0$ are asymptotically locally flat. In the case $A=0$ the black holes are solutions of the theory that only contains the $K$ term in the Lagrangian. Additionally, for $A>0$ and $B=0$ this
metric reduces to the non-rotating BTZ black hole.

\section{Dirac Quasinormal modes of New Type Black Holes}

The QNMs of fermionic perturbations in the background of new type black hole
are given by the fermionic field solution to the Dirac equation in curved
spacetime with suitable boundary conditions. This means that there is only
ingoing waves on the event horizon and we will consider that the fermionic field
vanishes at spatial infinity for asymptotically AdS black holes, known as
Dirichlet boundary conditions, and that there are only outgoing waves at the
cosmological horizon or at spatial infinity for asymptotically dS black
holes and for asymptotically locally flat black holes, respectively.
The Dirac equation is given by%
\begin{equation}
\left( \gamma ^{\mu }\nabla _{\mu }+m\right) \psi =0~,
\end{equation}%
where the covariant derivative is defined as 
\begin{equation}
\nabla _{\mu }=\partial _{\mu }+\frac{1}{2}\omega _{\text{ \ \ \ }\mu
}^{ab}J_{ab}~,
\end{equation}%
and the generators of the Lorentz group $J_{ab}$ are 
\begin{equation}
J_{ab}=\frac{1}{4}\left[ \gamma _{a},\gamma _{b}\right] ~.
\end{equation}%
The gamma matrices in curved spacetime $\gamma ^{\mu }$ are defined by 
\begin{equation}
\gamma ^{\mu }=e_{\text{ \ }a}^{\mu }\gamma ^{a}~,
\end{equation}%
where $\gamma ^{a}$ are the gamma matrices in flat spacetime. In order to
solve the Dirac equation we use the diagonal vielbein 
\begin{equation}
e^{0}=\sqrt{f \left(r \right)} dt~,\text{ \ }e^{1}=\frac{dr}{\sqrt{f \left(r
\right)}} ~,\text{ \ }e^{2}=rd\phi~,
\end{equation}
and, from the null torsion condition 
\begin{equation}
de^{a}+\omega _{\text{ \ }b}^{a}e^{b}=0~,
\end{equation}%
we obtain the spin connection 
\begin{equation}
\omega ^{01}=\frac{f^{\prime }\left(r \right)}{2}dt~,\text{ \ }\omega ^{12}=-%
\sqrt{f \left( r \right)} d\phi~.
\end{equation}%
Now, using the following representation of the gamma matrices 
\begin{equation}
\gamma ^{0}=i\sigma ^{2}~,\text{ \ }\gamma ^{1}=\sigma ^{1}~,\text{ \ }%
\gamma ^{2}=\sigma ^{3}~,
\end{equation}%
where $\sigma ^{i}$ are the Pauli matrices, and using the following ansatz
for the fermionic field 
\begin{equation}
\psi =\frac{e^{-i\omega t} e^{i\kappa \phi}}{\sqrt{f \left( r \right)^{1/2} r%
} }\left( 
\begin{array}{c}
\psi _{1} \\ 
\psi _{2}%
\end{array}%
\right) ~,
\end{equation}%
we arrive at the following coupled system of differential equations 
\begin{eqnarray}
\sqrt{f(r)}\psi _{1}^{\prime }+\frac{i\omega}{\sqrt{f(r)}}\psi _{1}-\frac{%
i\kappa }{r}\psi _{2}+m\psi _{2} &=&0~,  \notag  \label{ecuacion} \\
\sqrt{f(r)}\psi _{2}^{\prime }-\frac{i\omega}{\sqrt{f(r)}}\psi _{2}+\frac{%
i\kappa }{r}\psi _{1}+m\psi _{1} &=&0~,
\end{eqnarray}
where $m$ is the mass of the fermionic field $\psi$, which is minimally
coupled to curvature. Decoupling the above system of equations we obtain the following equation for $\psi_{1}$
\begin{gather}
-2r^{2}\left( mr-i\kappa \right) f\left( r\right) ^{2}\psi _{1}^{\prime
\prime }+irf\left( r\right) \left( 2\kappa f\left( r\right) +r\left( \kappa
+imr\right) f^{\prime }\left( r\right) \right) \psi _{1}^{\prime }+  \notag
\\
\left( 2\left( m^{3}r^{3}-im^{2}r^{2}\kappa +mr\kappa ^{2}-i\kappa
^{3}-\kappa \omega r\right) f\left( r\right) -r^{2}\left( mr-i\kappa \right)
\omega \left( 2\omega -if^{\prime }\left( r\right) \right) \right) \psi
_{1}=0~.  \label{radial}
\end{gather}
Now, in order to obtain analytical solutions we will consider some special
cases.

\subsection{Null angular momentum}

In this section we will consider the case $\kappa =0$. So, making the change
of variables $y=1-\frac{r_{+}}{r}$, the equation (\ref{radial}) becomes 
\begin{gather}
\psi _{1}^{\prime \prime }\left( y\right) +\left( \frac{1/2}{y}+\frac{1}{y-1}%
+\frac{1/2}{y-1+Q}\right) \psi _{1}^{\prime }\left( z\right) +  \notag \\
\left( -\frac{m^{2}Q/A}{y-1}+\frac{\frac{\omega ^{2}Q^{2}}{\left( 1-Q\right)
A^{2}r_{+}^{2}}+\frac{i\omega Q}{2Ar_{+}}}{y}+\frac{-\frac{\omega ^{2}Q^{3}}{%
\left( 1-Q\right) A^{2}r_{+}^{2}}+\frac{i\omega Q^{2}}{2Ar_{+}}}{y-1+Q}%
\right) \frac{1}{y\left( y-1\right) \left( y-1+Q\right) }\psi _{1}\left(
y\right) =0~.  \label{diff}
\end{gather}
We note that this equation correspond to a Riemann's differential equation,
whose general form is \cite{M. Abramowitz} 
\begin{eqnarray}  \label{eqn}
&&\frac{d^{2}w}{dz^{2}}+\left( \frac{1-\alpha -\alpha ^{\prime }}{z-r}+\frac{%
1-\beta -\beta ^{\prime }}{z-s}+\frac{1-\gamma -\gamma ^{\prime }}{z-t}%
\right) \frac{dw}{dz}+  \notag \\
&&\left( \frac{\alpha \alpha ^{\prime }\left( r-s\right) \left( r-t\right) }{%
z-r}+\frac{\beta \beta ^{\prime }\left( s-t\right) \left( s-r\right) }{z-s}+%
\frac{\gamma \gamma ^{\prime }\left( t-r\right) \left( t-s\right) }{z-t}%
\right) \frac{w}{\left( z-r\right) \left( z-s\right) \left( z-t\right) }=0~,
\end{eqnarray}%
where $r,s,t$ are the singular points, and the exponents $\alpha,\alpha
^{\prime }, \beta, \beta^{\prime}, \gamma, \gamma^{\prime}$ are subject to
the condition 
\begin{equation}
\alpha +\alpha ^{\prime }+\beta +\beta ^{\prime }+\gamma +\gamma ^{\prime
}=1~.
\end{equation}%
The complete solution of (\ref{eqn}) is denoted by the symbol 
\begin{equation}
w=P\left\{ 
\begin{array}{cccc}
r & s & t &  \\ 
\alpha & \beta & \gamma & z \\ 
\alpha ^{\prime } & \beta ^{\prime } & \gamma ^{\prime } & 
\end{array}%
\right\} ~,
\end{equation}%
where the $P$ symbol denotes the Riemann's $P$ function, which can be
reduced to the hypergeometric function through 
\begin{equation}
w=\left( \frac{z-r}{z-s}\right) ^{\alpha }\left( \frac{z-t}{z-s}\right)
^{\gamma }P\left\{ 
\begin{array}{cccc}
0 & \infty & 1 &  \\ 
0 & \alpha +\beta +\gamma & 0 & \frac{\left( z-r\right) \left( t-s\right) }{%
\left( z-s\right) \left( t-r\right) } \\ 
\alpha ^{\prime }-\alpha & \alpha +\beta ^{\prime }+\gamma & \gamma ^{\prime
}-\gamma & 
\end{array}%
\right\} ~,
\end{equation}%
where  the $P$ function is now the Gauss' hypergeometric function.

\subsubsection{Asymptotically $AdS$ New Type Black Holes}

In this case $r_{+}>r_{-}$ ($A>0$). So, considering equations (\ref{diff}) and (\ref{eqn}) we can identify the regular singular points $r, s$ and $t$ as 
\begin{equation}
r=0~,\text{ \ }s=1-Q~,\text{ \ }t=1~.
\end{equation}%
Therefore, the exponents are given by 
\begin{eqnarray}
\alpha &=&-\frac{i\omega Q}{Ar_{+}\left( Q-1\right) }~,\ \ \alpha ^{\prime } =\frac{1}{2}+\frac{i\omega Q}{Ar_{+}\left( Q-1\right) }~,\\
\beta &=&\frac{i\omega Q}{Ar_{+}\left( Q-1\right) }~, \ \ \ \ \ \beta ^{\prime } =\frac{1}{2}-\frac{i\omega Q}{Ar_{+}\left( Q-1\right) }~,
\\
\gamma &=&\frac{m}{\sqrt{A}}~, \ \ \ \ \ \ \ \ \ \  \ \ \ \ \ \ \gamma ^{\prime } =-\frac{m}{\sqrt{A}}~,
\end{eqnarray}%
and the solution to equation (\ref{diff}) can be written as 
\begin{eqnarray}
\psi _{1}\left( y\right) &=&C_{1}\left( \frac{y}{y-1+Q}\right) ^{\alpha
}\left( \frac{y-1}{y-1+Q}\right) ^{\gamma }{_{2}}F{_{1}}\left( a,b,c,\frac{Qy%
}{y-(1-Q)}\right) +  \notag \\
&&C_{2}\left( \frac{y}{y-1+Q}\right) ^{\alpha ^{\prime }}\left( \frac{y-1}{%
y-1+Q}\right) ^{\gamma }{_{2}}F{_{1}}\left( a-c+1,b-c+1,2-c,\frac{Qy}{y-(1-Q)%
}\right) ~,
\end{eqnarray}%
where we have defined the constants $a, b$ and $c$ as 
\begin{eqnarray}
a &=&\alpha +\beta +\gamma ~,  \notag \\
b &=&\alpha +\beta ^{\prime }+\gamma ~,  \notag \\
c &=&1+\alpha -\alpha ^{\prime }~.
\end{eqnarray}%
In the near horizon limit, the above expression behaves as 
\begin{equation}
\psi _{1}\left( y\rightarrow 0\right) =\widehat{C}_{1}y^{\alpha }+\widehat{C}%
_{2}y^{\alpha ^{\prime }}~.
\end{equation}%
Now, we impose as boundary condition that classically nothing can escape
from the event horizon. Note that $Q/A\left( Q-1\right) >0$,
therefore we must take $C_{2}=0$ in order to have only ingoing waves at the
horizon. So, the solution simplifies to 
\begin{equation}
\psi _{1}\left( y\right) =C_{1}\left( \frac{y}{y-1+Q}\right) ^{\alpha
}\left( \frac{y-1}{y-1+Q}\right) ^{\gamma }{_{2}}F{_{1}}\left( a,b,c,\frac{Qy%
}{y-(1-Q)}\right) ~.
\end{equation}%
Now, we implement boundary conditions at spatial infinity $y\rightarrow 1$.
In order to do this, we employ the Kummer's relations \cite{M. Abramowitz},
which allow us to write the solution as 
\begin{eqnarray}
\psi _{1}\left( y\right) &=&C_{1}\left( \frac{y}{y-1+Q}\right) ^{\alpha
}\left( \frac{y-1}{y-1+Q}\right) ^{\gamma }\frac{\Gamma \left( c\right)
\Gamma \left( c-a-b\right) }{\Gamma \left( c-a\right) \Gamma \left(
c-b\right) }{_{2}}F{_{1}}\left( a,b,a+b-c,1-\frac{Qy}{y-\left( 1-Q\right) }%
\right) +  \notag \\
&&C_{1}\left( 1-Q\right) ^{\gamma ^{\prime }-\gamma }\left( \frac{y}{y-1+Q}%
\right) ^{\alpha }\left( \frac{y-1}{y-1+Q}\right) ^{\gamma ^{\prime }}\frac{%
\Gamma \left( c\right) \Gamma \left( a+b-c\right) }{\Gamma \left( a\right)
\Gamma \left( b\right) }\times  \notag \\
&&{_{2}}F{_{1}}\left( c-a,c-b,c-a-b+1,1-\frac{Qy}{y-\left( 1-Q\right) }%
\right) ~.
\end{eqnarray}%
In the limit $y\rightarrow 1$, the above expression becomes 
\begin{equation}
\psi _{1}\left( y\rightarrow 1\right) =\widetilde{C}_{1}\left( 1-y\right)
^{\gamma }\frac{\Gamma \left( c\right) \Gamma \left( c-a-b\right) }{\Gamma
\left( c-a\right) \Gamma \left( c-b\right) }+\widetilde{C}_{1}\left(
1-Q\right) ^{\gamma ^{\prime }-\gamma }\left( 1-y\right) ^{\gamma ^{\prime }}%
\frac{\Gamma \left( c\right) \Gamma \left( a+b-c\right) }{\Gamma \left(
a\right) \Gamma \left( b\right) }~.  \label{infinity}
\end{equation}%
So, imposing that the scalar field be null at spatial infinity, we can
determine the quasinormal frequencies. For $m>0$ the second term of equation
(\ref{infinity}) blows up when $y\rightarrow 1$ unless we impose the
condition $a=-n$ or $b=-n$, these conditions give the following QNFs 
\begin{equation}
\omega =-i\frac{A\left( r_{+}-r_{-}\right) }{4}\left( 1+2n+\frac{2m}{\sqrt{A}%
}\right)~.  \label{omegas}
\end{equation}
Due to the imaginary part of the QNFs is negative the asymptotically $AdS$
new type black holes are stable under fermionic perturbations, at least for
the mode with the lowest angular momentum. Also, in a similar way the
quasinormal frequencies associated to $\psi _{2}$ can be obtained, note that 
$\psi _{2}$ satisfies a similar equation that $\psi _{1}$ but making the
changes $\kappa \rightarrow -\kappa $ and $\omega \rightarrow -\omega $.

\subsubsection{Asymptotically $dS$ New Type Black Holes}

In this case, besides the event horizon $r_{+}$, we have a cosmological
horizon $r_{-}$, where $r_{-}>r_{+}$ ($A<0$). So, considering equations (\ref{diff}) and (\ref{eqn}) we can identify the regular singular points $r, s$ and $t$ as 
\begin{equation}
r=0~,\text{ \ }s=1~,\text{ \ }t=1-Q~.
\end{equation}%
Therefore, the exponents are given by 
\begin{eqnarray}
\alpha &=&-\frac{i\omega Q}{Ar_{+}\left( Q-1\right) }~, \ \ \alpha ^{\prime } =\frac{1}{2}+\frac{i\omega Q}{Ar_{+}\left( Q-1\right) }~,
\\
\beta &=&\frac{m}{\sqrt{A}}~, \ \ \ \ \ \ \ \ \ \ \ \ \ \ \ \ \beta ^{\prime } =-\frac{m}{\sqrt{A}}~, \\
\gamma &=&\frac{i\omega Q}{Ar_{+}\left( Q-1\right) }~, \ \ \ \ \ \gamma ^{\prime } =\frac{1}{2}-\frac{i\omega Q}{Ar_{+}\left( Q-1\right) }~,
\end{eqnarray}
and the solution to equation (\ref{diff}) can be written as 
\begin{eqnarray}
\psi _{1}\left( y\right) &=&C_{1}\left( \frac{y}{y-1}\right) ^{\alpha
}\left( \frac{y-1+Q}{y-1}\right) ^{\gamma }{_{2}}F{_{1}}\left( a,b,c,-\frac{%
Qy}{\left( y-1\right) (1-Q)}\right) +  \notag \\
&&C_{2}\left( \frac{y}{y-1}\right) ^{\alpha ^{\prime }}\left( \frac{y-1+Q}{%
y-1}\right) ^{\gamma }{_{2}}F{_{1}}\left( a-c+1,b-c+1,2-c,-\frac{Qy}{\left(
y-1\right) (1-Q)}\right) ~,
\end{eqnarray}%
where we have defined the constants $a, b$ and $c$ as 
\begin{eqnarray}
a &=&\alpha +\beta +\gamma ~,  \notag \\
b &=&\alpha +\beta ^{\prime }+\gamma ~,  \notag \\
c &=&1+\alpha -\alpha ^{\prime }~.
\end{eqnarray}%
In the near horizon limit, the above expression behaves as 
\begin{equation}
\psi _{1}\left( y\rightarrow 0\right) =\widehat{C}_{1}y^{\alpha }+\widehat{C}%
_{2}y^{\alpha ^{\prime}}~.
\end{equation}%
Now, we impose as boundary condition that classically nothing can escape
from the event horizon. Note that $Q/A\left( Q-1\right) >0$ as in the case $1$,
therefore we must take $C_{2}=0$ in order to have only ingoing waves at the
horizon. So, the solution simplifies to 
\begin{equation}
\psi _{1}\left( y\right) =C_{1}\left( \frac{y}{y-1}\right) ^{\alpha }\left( 
\frac{y-1+Q}{y-1}\right) ^{\gamma }{_{2}}F{_{1}}\left( a,b,c,-\frac{Qy}{%
\left( y-1\right) (1-Q)}\right) ~.
\end{equation}%
Now, we implement boundary conditions at the cosmological horizon $%
y\rightarrow 1-Q$. In order to do this, we employ the Kummer's relations 
\cite{M. Abramowitz}, which allow us to write the solution as 
\begin{eqnarray}
\psi _{1}\left( y\right) &=&C_{1}\left( \frac{y}{y-1}\right) ^{\alpha
}\left( \frac{y-1+Q}{y-1}\right) ^{\gamma }\frac{\Gamma \left( c\right)
\Gamma \left( c-a-b\right) }{\Gamma \left( c-a\right) \Gamma \left(
c-b\right) }{_{2}}F{_{1}}\left( a,b,a+b-c,1+\frac{Qy}{\left( y-1\right) (1-Q)%
}\right) +  \notag \\
&&C_{1}\left( \frac{1}{1-Q}\right) ^{\gamma ^{\prime }-\gamma }\left( \frac{y%
}{y-1}\right) ^{\alpha }\left( \frac{y-1+Q}{y-1}\right) ^{\gamma ^{\prime }}%
\frac{\Gamma \left( c\right) \Gamma \left( a+b-c\right) }{\Gamma \left(
a\right) \Gamma \left( b\right) }\times  \notag \\
&&{_{2}}F{_{1}}\left( c-a,c-b,c-a-b+1,1+\frac{Qy}{\left( y-1\right) (1-Q)}%
\right)~.
\end{eqnarray}%
In the limit $y\rightarrow 1-Q$, the above expression becomes 
\begin{eqnarray}
\psi _{1}\left( y\rightarrow 1\right) &=&\widetilde{C}_{1}\left(
1-Q-y\right) ^{\gamma }\frac{\Gamma \left( c\right) \Gamma \left(
c-a-b\right) }{\Gamma \left( c-a\right) \Gamma \left( c-b\right) }+%
\widetilde{C}_{1}\left( \frac{1}{1-Q}\right) ^{\gamma ^{\prime }-\gamma
}\left( 1-Q-y\right) ^{\gamma ^{\prime }}\frac{\Gamma \left( c\right) \Gamma
\left( a+b-c\right) }{\Gamma \left( a\right) \Gamma \left( b\right) }= 
\notag \\
&&~\widetilde{C}_{1}e^{-\gamma \ln \left( \frac{1}{1-Q-y}\right) }\frac{%
\Gamma \left( c\right) \Gamma \left( c-a-b\right) }{\Gamma \left( c-a\right)
\Gamma \left( c-b\right) }+\widetilde{C}_{1}\left( \frac{1}{1-Q}\right)
^{\gamma ^{\prime }-\gamma }e^{-\left( 1/2-\gamma \right) \ln \left( \frac{1%
}{1-Q-y}\right) }\frac{\Gamma \left( c\right) \Gamma \left( a+b-c\right) }{%
\Gamma \left( a\right) \Gamma \left( b\right)}~.
\end{eqnarray}
So, in order to have only outgoing waves at the cosmological horizon, we must
impose $c-a=-n$ or $c-b=-n$. These conditions yield the following set of
quasinormal frequencies 
\begin{equation}
\omega =\pm \frac{\left( r_{-}-r_{+}\right) \sqrt{\left\vert A\right\vert }m%
}{2}-i\frac{\left( r_{-}-r_{+}\right) \left\vert A\right\vert }{4}\left(
1+2n\right)~.
\end{equation}
Due to the imaginary part of the QNFs is negative the asymptotically $dS$
new type black holes are stable under fermionic perturbations, at least for
the mode with the lowest angular momentum. As in the previous case, in a
similar way the quasinormal frequencies associated to $\psi _{2}$ can be
obtained, note that $\psi _{2}$ satisfies a similar equation that $\psi _{1} 
$ but making the changes $\kappa \rightarrow -\kappa $ and $\omega
\rightarrow -\omega $.

\subsection{Massless Dirac QNMs of Asymptotically $dS$ and Asymptotically
Locally Flat New Type Black Holes}

In this section we compute the QNMs for asymptotically $dS$ ($A<0$) and
asymptotically locally flat ($A=0$) new type black holes  for massless
fermionic perturbations ($m=0$).

\subsubsection{Asymptotically $dS$ New Type Black Holes}

Performing the change of variables $y=1-\frac{r_{+}}{r}$, the equation (\ref%
{radial}) reduces to 
\begin{equation}
y\left( y-1+Q\right) \psi _{1}^{\prime \prime }\left( y\right) +\frac{1}{2}%
\left( 2y-1+Q\right) \psi _{1}^{\prime }\left( y\right) +\left( -\frac{%
\kappa ^{2}Q}{Ar_{+}^{2}}+\frac{\omega ^{2}Q^{2}}{A^{2}r_{+}^{2}y\left(
y-1+Q\right) }-\frac{i\omega Q}{2Ar_{+}}\left( \frac{1}{y-1+Q}+\frac{1}{y}%
\right) \right) \psi_{1} =0~,
\end{equation}%
where we have defined $Q=r_{+}/r_{-}<1$. It is worth mentioning that in
these coordinates the event horizon is located at $y=0$, and the
cosmological horizon at $y=1-Q$. Now, making the change of variables 
\begin{equation}
\psi _{1}^{\prime \prime }\left( y\right) =y^{\alpha }\left( y-1+Q\right)
^{\beta }F\left( y\right) ~,
\end{equation}%
we arrive at the following equation 
\begin{equation}
y\left( y-1+Q\right) F^{\prime \prime }\left( y\right) +\frac{1}{2}\left(
\left( Q-1\right) \left( 1+4\alpha \right) +2y\left( 1+2\alpha +2\beta
\right) \right) F^{\prime }\left( y\right) +\left( -\frac{\kappa ^{2}Q}{%
Ar_{+}^{2}}+\left( \alpha +\beta \right) ^{2}\right) F\left( y\right) =0~,
\end{equation}%
where $\alpha $ and $\beta $ are given by 
\begin{eqnarray}
\alpha _{+} &=&\frac{iQ\omega }{Ar_{+}\left( 1-Q\right) }~,\text{ \ }\alpha
_{-}=\frac{1}{2}-\frac{iQ\omega }{Ar_{+}\left( 1-Q\right) }~,  \notag \\
\beta _{+} &=&-\frac{iQ\omega }{Ar_{+}\left( 1-Q\right) }~,\text{ \ }\beta
_{-}=\frac{1}{2}+\frac{iQ\omega }{Ar_{+}\left( 1-Q\right) }~.
\end{eqnarray}%
In the next, we will use the values $\alpha =\frac{iQ\omega }{Ar_{+}\left(
1-Q\right) }$ and $\beta =-\frac{iQ\omega }{Ar_{+}\left( 1-Q\right) }$.
Then, performing another change of variable $\left( 1-Q\right) z=y$ we
obtain 
\begin{equation}
z\left( 1-z\right) F^{\prime \prime }\left( z\right) +\left( \frac{1}{2}%
+2\alpha -z\left( 1+2\alpha +2\beta \right) \right) F^{\prime }\left(
z\right) +\left( \frac{\kappa ^{2}Q}{Ar_{+}^{2}}-\left( \alpha +\beta
\right) ^{2}\right) F\left( z\right) =0~,  \label{hipergeometrica}
\end{equation}%
and, in these new coordinates the event horizon is located at $z=0$, and the
cosmological horizon at $z=1$. We recognize equation (\ref%
{hipergeometrica}) as the hypergeometric equation 
\begin{equation}
z\left( 1-z\right) F^{\prime \prime }\left( z\right) +\left( c-\left(
1+a+b\right) z\right) F^{\prime }\left( z\right) -abF\left( z\right) =0~,
\end{equation}%
where 
\begin{eqnarray}
a &=&\alpha +\beta \pm \frac{i\kappa \sqrt{Q}}{\sqrt{\left\vert A\right\vert 
}r_{+}}~,  \notag \\
b &=&\alpha +\beta \mp \frac{i\kappa \sqrt{Q}}{\sqrt{\left\vert A\right\vert 
}r_{+}}~,  \notag \\
c &=&\frac{1}{2}+2\alpha ~.
\end{eqnarray}%
The general solution of the hypergeometric equation is 
\begin{equation}
F\left( z\right) =C_{1}\ {_{2}}F_{1}\left( a,b,c;z\right) +C_{2}z^{1-c}{_{2}}%
F_{1}\left( a-c+1,b-c+1,2-c;z\right) ~,
\end{equation}%
which has three regular singular points at $z=0$, $z=1$ and $z=\infty $.
Here, $_{2}F_{1}(a,b,c;z)$ is a hypergeometric function and $C_{1}$, $C_{2}$
are constants. Then, the solution for the radial function $\psi _{1}\left(
z\right) $ is 
\begin{equation}
\psi _{1}\left( z\right) =C_{1}z^{\alpha }\left( 1-z\right) ^{\beta }{_{2}}%
F_{1}\left( a,b,c;z\right) +C_{2}z^{1/2-\alpha }\left( 1-z\right) ^{\beta }{%
_{2}}F_{1}\left( a-c+1,b-c+1,2-c;z\right) ~.
\end{equation}%
So, in the vicinity of the event horizon $z=0$, and using the property ${_{2}%
}F_{1}\left( a,b,c;0\right) =1$, the function $\psi _{1}\left( z\right) $
behaves as 
\begin{equation}
\psi _{1}\left( z\right) =C_{1}e^{\alpha \ln z}+C_{2}e^{\left( 1/2-\alpha
\right) \ln z}~.
\end{equation}%
Now, imposing boundary conditions at the event horizon, that there is only ingoing
modes, implies that $C_{2}=0$. Thus, the solution can be written as 
\begin{equation}
\psi _{1}\left( z\right) =C_{1}z^{\alpha }\left( 1-z\right) ^{\beta }{_{2}}%
F_{1}\left( a,b,c;z\right) ~.
\end{equation}%
On the other hand, using Kummer's formula for hypergeometric functions, 
\begin{gather}
{_{2}}F_{1}\left( a,b,c;z\right) =\frac{\Gamma \left( c\right) \Gamma \left(
c-a-b\right) }{\Gamma \left( c-a\right) \Gamma \left( c-b\right) }{_{2}}%
F_{1}\left( a,b,a+b-c;1-z\right) +  \notag \\
\left( 1-z\right) ^{c-a-b}\frac{\Gamma \left( c\right) \Gamma \left(
a+b-c\right) }{\Gamma \left( a\right) \Gamma \left( b\right) }{_{2}}%
F_{1}\left( c-a,c-b,c-a-b+1;1-z\right) ~,
\label{relationkummer}
\end{gather}%
the radial function $\psi _{1}\left( z\right) $
can be written as 
\begin{eqnarray}
\psi _{1}\left( z\rightarrow 1\right)  &=&C_{1}\left( 1-z\right) ^{\beta }%
\frac{\Gamma \left( c\right) \Gamma \left( c-a-b\right) }{\Gamma \left(
c-a\right) \Gamma \left( c-b\right) }+C_{1}\left( 1-z\right) ^{1/2-\beta }%
\frac{\Gamma \left( c\right) \Gamma \left( a+b-c\right) }{\Gamma \left(
a\right) \Gamma \left( b\right) }  \notag \\
&=&C_{1}e^{-\beta \ln \left( \frac{1}{1-z}\right) }\frac{\Gamma \left(
c\right) \Gamma \left( c-a-b\right) }{\Gamma \left( c-a\right) \Gamma \left(
c-b\right) }+C_{1}e^{\left( \beta -1/2\right) \ln \left( \frac{1}{1-z}%
\right) }\frac{\Gamma \left( c\right) \Gamma \left( a+b-c\right) }{\Gamma
\left( a\right) \Gamma \left( b\right) }~.
\end{eqnarray}%
So, in order to have only outgoing waves at the cosmological horizon $z=1$, we must
impose $c-a=-n$ or $c-b=-n$. These conditions yield the following set of
quasinormal frequencies 
\begin{equation}
\omega =\pm \frac{\left( 1-Q\right) \sqrt{\left\vert A\right\vert }\kappa }{2%
\sqrt{Q}}-i\frac{\left( 1-Q\right) \left\vert A\right\vert r_{+}}{4Q}\left(
1+2n\right) ~.
\end{equation}%
In a similar way the quasinormal frequencies associated to $\psi _{2}$ can
be obtained, note that $\psi _{2}$ satisfies a similar equation that $\psi
_{1}$ but making the changes $\kappa \rightarrow -\kappa $ and $\omega
\rightarrow -\omega $.

\subsubsection{Asymptotically Locally Flat New Type Black Holes}

Under the change of variables $y=1-\frac{r_{+}}{r}$, the equation (%
\ref{radial}) becomes 
\begin{equation}
y\left( y-1\right) \psi _{1}^{\prime \prime }\left( y\right) +\frac{1}{2}%
\left( 2y-1\right) \psi _{1}^{\prime }\left( y\right) +\left( \frac{\kappa
^{2}}{Br_{+}}+\frac{\omega ^{2}}{B^{2}y\left( y-1\right) }+\frac{i\omega }{2B%
}\left( \frac{1}{y-1}+\frac{1}{y}\right) \right) =0~,
\end{equation}%
Now, making the change of variables 
\begin{equation}
\psi _{1}^{\prime \prime }\left( y\right) =y^{\alpha }\left( y-1\right)
^{\beta }F\left( y\right) ~,
\end{equation}%
we arrive at the following equation 
\begin{equation}\label{hipergeometricaF}
y\left( 1-y\right) F^{\prime \prime }\left( y\right) +\frac{1}{2}\left(
1+4\alpha -2y\left( 1+2\alpha +2\beta \right) \right) F^{\prime }\left(
y\right) +\left( -\frac{\kappa ^{2}}{Br_{+}}-\left( \alpha +\beta \right)
^{2}\right) F\left( y\right) =0~,
\end{equation}%
where $\alpha $ and $\beta $ are given by 
\begin{eqnarray}
\alpha _{+} &=&\frac{1}{2}+\frac{i\omega }{B}~,\text{ \ }\alpha _{-}=-\frac{%
i\omega }{B}~,  \notag \\
\beta _{+} &=&\frac{i\omega }{B}~,\text{ \ }\beta _{-}=\frac{1}{2}-\frac{%
i\omega }{B}~.
\end{eqnarray}%
In the next, we will use the values $\alpha =-\frac{i\omega }{B}$ and $\beta
=\frac{i\omega }{B}$. We recognize equation (\ref{hipergeometricaF}) as the
hypergeometric equation 
\begin{equation}
y\left( 1-y\right) F^{\prime \prime }\left( y\right) +\left( c-\left(
1+a+b\right) y\right) F^{\prime }\left( y\right) -abF\left( y\right) =0~,
\end{equation}%
where 
\begin{eqnarray}
a &=&\alpha +\beta \pm \frac{i\kappa }{\sqrt{Br_{+}}}~,  \notag \\
b &=&\alpha +\beta \mp \frac{i\kappa }{\sqrt{Br_{+}}}~,  \notag \\
c &=&\frac{1}{2}+2\alpha ~.
\end{eqnarray}%
As in the previous case, the general solution of the hypergeometric equation is 
\begin{equation}
F\left( y\right) =C_{1}\ {_{2}}F_{1}\left( a,b,c;y\right) +C_{2}y^{1-c}{_{2}}%
F_{1}\left( a-c+1,b-c+1,2-c;y\right) ~,
\end{equation}%
which has three regular singular points at $y=0$, $y=1$ and $y=\infty $.
Here, $_{2}F_{1}(a,b,c;y)$ is a hypergeometric function and $C_{1}$, $C_{2}$
are constants. Then, the solution for the radial function $\psi _{1}\left(
y\right) $ is 
\begin{equation}
\psi _{1}\left( y\right) =C_{1}y^{\alpha }\left( 1-y\right) ^{\beta }{_{2}}%
F_{1}\left( a,b,c;y\right) +C_{2}y^{1/2-\alpha }\left( 1-y\right) ^{\beta }{%
_{2}}F_{1}\left( a-c+1,b-c+1,2-c;y\right) ~.
\end{equation}%
So, in the vicinity of the event horizon, $y=0$ and using the property ${_{2}%
}F_{1}\left( a,b,c;0\right) =1$, the function $\psi _{1}\left( y\right) $
behaves as 
\begin{equation}
\psi _{1}\left( z\right) =C_{1}e^{\alpha \ln y}+C_{2}e^{\left( 1/2-\alpha
\right) \ln y}~.
\end{equation}%
Now, imposing boundary conditions at the horizon, that there is only ingoing
modes, implies that $C_{2}=0$. Thus, the solution can be written as 
\begin{equation}
\psi _{1}\left( y\right) =C_{1}z^{\alpha }\left( 1-y\right) ^{\beta }{_{2}}%
F_{1}\left( a,b,c;y\right) ~.
\end{equation}%
On the other hand, using Kummer's formula for hypergeometric functions (\ref{relationkummer}), 
the radial function $\psi _{1}\left( y\right) $ can
be written as 
\begin{eqnarray}
\psi _{1}\left( y\rightarrow 1\right)  &=&C_{1}\left( 1-y\right) ^{\beta }%
\frac{\Gamma \left( c\right) \Gamma \left( c-a-b\right) }{\Gamma \left(
c-a\right) \Gamma \left( c-b\right) }+C_{1}\left( 1-y\right) ^{1/2-\beta }%
\frac{\Gamma \left( c\right) \Gamma \left( a+b-c\right) }{\Gamma \left(
a\right) \Gamma \left( b\right) }  \notag \\
&=&C_{1}e^{-\beta \ln \left( \frac{1}{1-y}\right) }\frac{\Gamma \left(
c\right) \Gamma \left( c-a-b\right) }{\Gamma \left( c-a\right) \Gamma \left(
c-b\right) }+C_{1}e^{\left( \beta -1/2\right) \ln \left( \frac{1}{1-y}%
\right) }\frac{\Gamma \left( c\right) \Gamma \left( a+b-c\right) }{\Gamma
\left( a\right) \Gamma \left( b\right) }~.
\end{eqnarray}%
So, in order to have only outgoing waves at the spatial infinity $y=1$, we must
impose $c-a=-n$ or $c-b=-n$. These conditions yield the following set of
quasinormal frequencies 
\begin{equation}
\omega =\pm \frac{\kappa }{2}\sqrt{\frac{B}{r_{+}}}-i\frac{B}{4}\left(
1+2n\right) ~.
\end{equation}

\subsection{Massless Dirac QNMs of Extremal New Type Black Holes}

In this section we consider the extremal case $r_{+}=r_{-}$ $(A>0)$. First,
we take $\kappa =0$. So, using the change of variable $y=1-\frac{r_{+}}{r}$
as in the previous sections, equation (\ref{radial}) reduces to 

\begin{equation*}
y\left( 1-y\right) \psi _{1}^{\prime \prime }\left( y\right) +\left(
1-2y\right) \psi _{1}^{\prime }\left( y\right) +\left( -\frac{m^{2}}{%
Ay\left( 1-y\right) }-\frac{i\omega }{Ar_{+}y^{2}}+\frac{\omega ^{2}\left(
1-y\right) }{A^{2}r_{+}^{2}y^{3}}\right) \psi _{1}\left( y\right) =0~.
\end{equation*}

The solution is given in terms of Whittaker functions

\begin{eqnarray*}
\psi _{1}\left( y\right)  &=&C_{1}\sqrt{\frac{y}{1-y}}WhittakerM\left( 
\frac{1}{2},\frac{m}{\sqrt{A}},-\frac{2i\omega \left( 1-y\right) }{Ar_{+}y}%
\right) + \\
&&C_{2}\sqrt{\frac{y}{1-y}}WhittakerW\left( \frac{1}{2},\frac{m}{\sqrt{A}}%
,-\frac{2i\omega \left( 1-y\right) }{Ar_{+}y}\right) .
\end{eqnarray*}

In order to have a regular scalar field at spatial infinity $y\rightarrow 1$%
, we must set $C_{2}=0$. Therefore, the solution reduces to

\begin{equation*}
\psi _{1}\left( y\right) =C_{1}\sqrt{\frac{y}{1-y}}WhittakerM\left( \frac{1%
}{2},\frac{m}{\sqrt{A}},-\frac{2i\omega \left( 1-y\right) }{Ar_{+}y}\right) .
\end{equation*}

However, for $y\rightarrow 1$ this expression becomes null, which
shows the absence of QNMs.
Furthermore, for $m=0$ we obtain the equation

\begin{equation}
y\left( 1-y\right) \psi _{1}^{\prime \prime }\left( y\right) +\left(
1-y\right) \psi _{1}^{\prime }\left( y\right) +\left( -\frac{\kappa
^{2}\left( 1-y\right) }{Ar_{+}^{2}y}+\frac{i\omega }{Ar_{+}y}-\frac{i\omega 
}{Ar_{+}y^{2}}+\frac{\omega ^{2}\left( 1-y\right) }{A^{2}r_{+}^{2}y^{3}}%
\right) \psi _{1}\left( y\right) =0~,
\end{equation}%
whose solution is given in terms of confluent Heun functions 
\begin{eqnarray}
\psi _{1}\left( y\right)  &=&C_{1}e^{i\omega /(A r_{+}y)}y^{\kappa /\sqrt{A}r_{+}}Heun_{C}\left( \frac{2i\omega }{ A r_{+}},-\frac{2\kappa }{\sqrt{A}r_{+}},-1,-\frac{i\omega }{%
Ar_{+}},\frac{i\omega }{Ar_{+}}+\frac{1}{2},\frac{1}{y}\right) +  \notag \\
&&C_{2}e^{i\omega /( A r_{+}y)}y^{-\kappa /\sqrt{A}%
r_{+}}Heun_{C}\left( \frac{2i\omega }{ A r_{+}},\frac{%
2\kappa }{\sqrt{A}r_{+}},-1,-\frac{i\omega }{Ar_{+}},\frac{i\omega }{Ar_{+}}+%
\frac{1}{2},\frac{1}{y}\right) ~.
\end{eqnarray}%
Then, at infinity $y\rightarrow 1$, the radial function vanishes. Therefore,
there are not quasinormal modes for extremal new type black holes.

\section{Conclusions}

In this work we have calculated analytically the QNMs of fermionic
perturbations for some special cases for new type black holes, which are solutions of
three-dimensional NMG, and also are solutions of conformal gravity in three
dimensions. The first case that we have analyzed are
massive fermionic fields perturbations without angular momentum ($\kappa=0$)
in the black hole background, and we have found the QNFs for asymptotically $%
AdS$ ($dS$) new type black holes. For asymptotically $AdS$ new type black
holes the QNFs are purely imaginary and negative, which ensures the
stability of the black hole under fermionic perturbations. For asymptotically 
$dS$ new type black holes the QNFs have a real and imaginary part, and
the imaginary part is negative, which ensures the stability of the black
hole under fermionic perturbations. It is worth mentioning that in these cases
the Dirac equation can be written as a Riemann differential equation, as in 
\cite{Catalan:2014ama}. Other cases, where is possible to find the QNFs
analytically is for asymptotically $dS$ and asymptotically locally flat new type black holes for massless
fermionic field perturbations, in these cases we have found that the QNFs have a
real and imaginary part, and the imaginary part is negative, which
ensures the stability of the black hole under fermionic perturbations.

Finally, we have analyzed fermionic field perturbations in the extremal new type black holes for some special cases, and we found that there are not
quasinormal modes as occurs, for instance, in \cite{Crisostomo:2004hj, Catalan:2014ama}, where the
authors have showed the absence of QNMs in the extremal BTZ black hole
and the extremal four-dimensional Lifshitz Black Hole in Conformal Gravity.
However, it was shown that it is possible to construct the QNMs of
three-dimensional extremal black holes in an algebraic way as the
descendents of the highest weight modes \cite{Chen:2010sn}, with hidden
conformal symmetry being an intrinsic property of the extremal black hole. Also,
it is worth to mention that the absence of QNMs for extremal black holes
does not always occur, for instance see \cite{Afshar:2010ii}, where the
authors have showed the presence of QNMs for the extremal BTZ black holes in
TMG.

\section*{Acknowledgments}

We thank to Julio Oliva for useful comments. This work was funded by the Comisi{\'o}n Nacional de Investigaci{\'o}n Cient{%
\'i}fica y Tecnol{\'o}gica through FONDECYT Grant 11121148 (Y.V.). P. A. G.
acknowledges the hospitality of the Universidad de La Serena where part of
this work was undertaken. \appendix


\begin{thebibliography}{99}


\bibitem{Banados:1992wn}  M.~Banados, C.~Teitelboim and J.~Zanelli,  
Phys.\ Rev.\ Lett.\ \textbf{69} (1992) 1849  [hep-th/9204099].  


\bibitem{Deser:1981wh}  S.~Deser, R.~Jackiw and S.~Templeton,  
Annals Phys.\ \textbf{140} (1982) 372  [Erratum-ibid.\ \textbf{185} (1988)
406]  [Annals Phys.\ \textbf{185} (1988) 406]  [Annals Phys.\ \textbf{281}
(2000) 409].  


\bibitem{Deser:1982vy}  S.~Deser, R.~Jackiw and S.~Templeton,  
Phys.\ Rev.\ Lett.\ \textbf{48} (1982) 975.  


\bibitem{Bergshoeff:2009hq}  E.~A.~Bergshoeff, O.~Hohm and P.~K.~Townsend,  
Phys.\ Rev.\ Lett.\ \textbf{102} (2009) 201301  [arXiv:0901.1766 [hep-th]].  


\bibitem{Bergshoeff:2009aq}  E.~A.~Bergshoeff, O.~Hohm and P.~K.~Townsend,  
Phys.\ Rev.\ D \textbf{79} (2009) 124042  [arXiv:0905.1259 [hep-th]].  


\bibitem{Bergshoeff:2009tb}  E.~A.~Bergshoeff, O.~Hohm and P.~K.~Townsend,  
Annals Phys.\ \textbf{325} (2010) 1118  [arXiv:0911.3061 [hep-th]].  


\bibitem{Bergshoeff:2009fj}  E.~Bergshoeff, O.~Hohm and P.~Townsend,  
J.\ Phys.\ Conf.\ Ser.\ \textbf{229} (2010) 012005  [arXiv:0912.2944
[hep-th]].  


\bibitem{Andringa:2009yc}  R.~Andringa, E.~A.~Bergshoeff, M.~de Roo,
O.~Hohm, E.~Sezgin and P.~K.~Townsend,  
Class.\ Quant.\ Grav.\ \textbf{27} (2010) 025010  [arXiv:0907.4658
[hep-th]].  


\bibitem{Bergshoeff:2010mf}  E.~A.~Bergshoeff, O.~Hohm, J.~Rosseel,
E.~Sezgin and P.~K.~Townsend,  
Class.\ Quant.\ Grav.\ \textbf{28} (2011) 015002  [arXiv:1005.3952
[hep-th]].  


\bibitem{Nakasone:2009bn}  M.~Nakasone and I.~Oda,  
Prog.\ Theor.\ Phys.\ \textbf{121} (2009) 1389  [arXiv:0902.3531 [hep-th]].  


\bibitem{Clement:2009gq}  G.~Clement,  
Class.\ Quant.\ Grav.\ \textbf{26} (2009) 105015  [arXiv:0902.4634
[hep-th]].  


\bibitem{AyonBeato:2009yq}  E.~Ayon-Beato, G.~Giribet and M.~Hassaine,  
JHEP \textbf{0905} (2009) 029  [arXiv:0904.0668 [hep-th]].  


\bibitem{Clement:2009ka}  G.~Clement,  
Class.\ Quant.\ Grav.\ \textbf{26} (2009) 165002  [arXiv:0905.0553
[hep-th]].  


\bibitem{AyonBeato:2009nh}  E.~Ayon-Beato, A.~Garbarz, G.~Giribet and
M.~Hassaine,  
Phys.\ Rev.\ D \textbf{80} (2009) 104029  [arXiv:0909.1347 [hep-th]].  

\bibitem{Oliva:2009ip}  J.~Oliva, D.~Tempo and R.~Troncoso,  
JHEP \textbf{0907}, 011 (2009)  [arXiv:0905.1545 [hep-th]].


\bibitem{Kim:2009jm}  W.~Kim and E.~J.~Son,  
Phys.\ Lett.\ B \textbf{678} (2009) 107  [arXiv:0904.4538 [hep-th]].  

\bibitem{Oda:2009ys}  I.~Oda,  
JHEP \textbf{0905} (2009) 064  [arXiv:0904.2833 [hep-th]].  

\bibitem{Liu:2009pha}  Y.~Liu and Y.~-W.~Sun,  
Phys.\ Rev.\ D \textbf{79} (2009) 126001  [arXiv:0904.0403 [hep-th]].  


\bibitem{Nakasone:2009vt}  M.~Nakasone and I.~Oda,  
Phys.\ Rev.\ D \textbf{79} (2009) 104012  [arXiv:0903.1459 [hep-th]].  

\bibitem{Deser:2009hb}  S.~Deser,  
Phys.\ Rev.\ Lett.\ \textbf{103} (2009) 101302  [arXiv:0904.4473 [hep-th]].  

\bibitem{Correa:2014ika}
  F.~Correa, M.~Hassaine and J.~Oliva,
  arXiv:1403.6479 [hep-th].


\bibitem{Regge:1957td}  T.~Regge and J.~A.~Wheeler,  
Phys.\ Rev.\ \textbf{108}, 1063 (1957).


\bibitem{Zerilli:1971wd}  F.~J.~Zerilli,  
Phys.\ Rev.\ D \textbf{2}, 2141 (1970).


\bibitem{Zerilli:1970se}  F.~J.~Zerilli,  
Phys.\ Rev.\ Lett.\ \textbf{24}, 737 (1970).  


\bibitem{Kokkotas:1999bd}  K.~D.~Kokkotas and B.~G.~Schmidt,  
Living Rev.\ Rel.\ \textbf{2}, 2 (1999)  [gr-qc/9909058].  


\bibitem{Nollert:1999ji}  H.~-P.~Nollert,  
Class.\ Quant.\ Grav.\ \textbf{16}, R159 (1999).  


\bibitem{Konoplya:2011qq}  R.~A.~Konoplya and A.~Zhidenko,  
Rev.\ Mod.\ Phys.\ \textbf{83}, 793 (2011)  [arXiv:1102.4014 [gr-qc]].

\bibitem{Chan:1996yk} 
  J.~S.~F.~Chan and R.~B.~Mann,
  Phys.\ Rev.\ D {\bf 55}, 7546 (1997)
  [gr-qc/9612026].

\bibitem{Cardoso:2001hn} 
  V.~Cardoso and J.~P.~S.~Lemos,
  Phys.\ Rev.\ D {\bf 63}, 124015 (2001)
  [gr-qc/0101052].

\bibitem{Birmingham:2001pj} 
  D.~Birmingham, I.~Sachs and S.~N.~Solodukhin,
  Phys.\ Rev.\ Lett.\  {\bf 88}, 151301 (2002)
  [hep-th/0112055].


\bibitem{Kwon:2011ey}  Y.~Kwon, S.~Nam, J.~-D.~Park and S.~-H.~Yi,  
Class.\ Quant.\ Grav.\ \textbf{28} (2011) 145006  [arXiv:1102.0138
[hep-th]].  


\bibitem{Maldacena:1997re}  J.~M.~Maldacena,  
Adv.\ Theor.\ Math.\ Phys.\ \textbf{2}, 231 (1998)  [hep-th/9711200].  


\bibitem{Horowitz:1999jd}  G.~T.~Horowitz and V.~E.~Hubeny,  
Phys.\ Rev.\ D \textbf{62}, 024027 (2000)  [hep-th/9909056].


\bibitem{Bekenstein:1974jk}  J.~D.~Bekenstein,  
Lett.\ Nuovo Cim.\ \textbf{11}, 467 (1974).  


\bibitem{Hod:1998vk}  S.~Hod,  
Phys.\ Rev.\ Lett.\ \textbf{81}, 4293 (1998)  [gr-qc/9812002].  


\bibitem{Kunstatter:2002pj}  G.~Kunstatter,  
Phys.\ Rev.\ Lett.\ \textbf{90}, 161301 (2003)  [gr-qc/0212014].


\bibitem{Maggiore:2007nq}  M.~Maggiore,  
Phys.\ Rev.\ Lett.\ \textbf{100}, 141301 (2008)  [arXiv:0711.3145 [gr-qc]].  

\bibitem{Corda:2012tz} 
  C.~Corda,
  Int.\ J.\ Mod.\ Phys.\ D {\bf 21}, 1242023 (2012)
  [arXiv:1205.5251 [gr-qc]].

\bibitem{Corda:2012dw} 
  C.~Corda,
  Eur.\ Phys.\ J.\ C {\bf 73}, 2665 (2013)
  [arXiv:1210.7747 [gr-qc]].

\bibitem{Corda:2013nza} 
  C.~Corda, S.~H.~Hendi, R.~Katebi and N.~O.~Schmidt,
  JHEP {\bf 1306}, 008 (2013)
  [arXiv:1305.3710 [gr-qc]].

\bibitem{Corda:2013paa} 
  C.~Corda, S.~H.~Hendi, R.~Katebi and N.~O.~Schmidt,
  arXiv:1401.2872 [physics.gen-ph].


\bibitem{Oliva:2009hz}  J.~Oliva, D.~Tempo and R.~Troncoso,  
Int.\ J.\ Mod.\ Phys.\ A \textbf{24} (2009) 1588  [arXiv:0905.1510
[hep-th]].  

\bibitem{M. Abramowitz} M. Abramowitz and A. Stegun, Handbook of
Mathematical functions, (Dover publications, New York, 1970).


\bibitem{Catalan:2014ama}  M.~Catalan, E.~Cisternas, P.~A.~Gonzalez and
Y.~Vasquez,  
arXiv:1404.3172 [gr-qc].  


\bibitem{Crisostomo:2004hj}  J.~Crisostomo, S.~Lepe and J.~Saavedra,  
Class.\ Quant.\ Grav.\ \textbf{21} (2004) 2801  [hep-th/0402048].  


\bibitem{Chen:2010sn}  B.~Chen and J.~-j.~Zhang,  
Phys.\ Lett.\ B \textbf{699} (2011) 204  [arXiv:1012.2219 [hep-th]].  


\bibitem{Afshar:2010ii}  H.~R.~Afshar, M.~Alishahiha and A.~E.~Mosaffa,  
JHEP \textbf{1008} (2010) 081  [arXiv:1006.4468 [hep-th]].  
\end{thebibliography}
\end{document}